# OPTIMAL FACTORIAL DESIGNS FOR CDNA MICROARRAY EXPERIMENTS

By Tathagata Banerjee and Rahul Mukerjee

*Indian Institute of Management Ahmedabad and Indian Institute of Management Calcutta*

We consider cDNA microarray experiments when the cell populations have a factorial structure, and investigate the problem of their optimal designing under a baseline parametrization where the objects of interest differ from those under the more common orthogonal parametrization. First, analytical results are given for the $2 \times 2$ factorial. Since practical applications often involve a more complex factorial structure, we next explore general factorials and obtain a collection of optimal designs in the saturated, that is, most economic, case. This, in turn, is seen to yield an approach for finding optimal or efficient designs in the practically more important nearly saturated cases. Thereafter, the findings are extended to the more intricate situation where the underlying model incorporates dye-coloring effects, and the role of dye-swapping is critically examined.

**1. Introduction.** Optimal designing of cDNA microarray experiments is an area of enormous potential that has started opening up in recent years. The fields of application include the biological, agricultural and pharmaceutical sciences. In an experimental design for microarrays, the cell populations under study represent the treatments which, as in traditional design theory, may be unstructured or have a factorial structure. The design is called varietal or factorial in these two situations respectively. In varietal designs, interest lies typically in all or some pairwise contrasts of the treatment effects, whereas in factorial designs, the objects of interest are the main effects of the factors and interactions among them. These factorial effects are commonly defined via an orthogonal parametrization, but a relatively less studied baseline parametrization, which is nonorthogonal, can also be of interest depending on the context. The main distinction between these two kinds of parametrization is that the former defines the factorial effects via mutually orthogonal treatment contrasts, whereas the latter defines these effects with reference to natural baseline levels of the factors and, hence, entails nonorthogonality. More details follow in Section 3.

In a pioneering paper Kerr and Churchill (2001a) discussed the design issues in microarrays and investigated optimal varietal designs that estimate the pairwise contrasts of treatment effects for fixed genes with minimum average variance. While observing that microarray designs can be considered as incomplete block designs with block size two, they noted the inadequacy of the classical optimality results in this regard and obtained, via complete enumeration, economic optimal designs for ten or fewer treatments. We refer to Kerr and Churchill (2001b), Yang and Speed (2002) and Churchill (2002) for very informative further discussion on the design issues. While Kerr and Churchill (2001a, 2001b) and Churchill (2002) concentrated on varietal designs, Yang and Speed (2002) discussed factorial designs in some detail. Subsequent work on varietal designs for microarrays includes those due to Dobbin and Simon (2002), Kerr (2003), Rosa, Steibel and Tempelman (2005), Wit, Nobile and Khanin (2005) and Altman and Hua (2006), although some of these authors, as also Churchill (2002), briefly touched upon factorial designs as well.

Work on factorial designs for microarrays started gaining momentum only very recently. A major reference in this regard is Kerr (2006), who worked under the framework of the orthogonal parametrization and explored two-level factorial designs that keep all main effects and two-factor interactions estimable, without any assumption on the absence of higher order interactions. Exploiting the connection between microarray designs and incomplete block designs with block size two, she showed how replicates arising from different blocking schemes can be combined for this purpose and, in particular, gave designs with the minimum number of replicates for eight or fewer factors. Related references in the general context of two-level factorials in blocks of size two include Yang and Draper (2003) and Wang (2004) and, interestingly, Kerr (2006) proposed a construction which is more economic than that in Wang (2004) for any number of factors. Bueno Filho, Gilmour and Rosa (2006) also considered factorial microarray designs for both fixed and random treatment effects. Their parametrization for fixed treatment effects is akin to the orthogonal one and, among other things, they studied optimal designs for the $3 \times 3$ factorial, paying more attention to the case where the two-factor interaction is absent. Further results on factorial microarray designs under the orthogonal parametrization were obtained by Landgrebe, Bretz and Brunner (2006), Gupta (2006) and Grossmann and Schwabe (2008). Landgrebe, Bretz and Brunner (2006) studied optimal designs within a collection of candidate designs and focused on the $2 \times 2$ and $3 \times 2$ factorials, while Gupta (2006) investigated the role of balanced factorial designs. Grossmann and Schwabe (2008) explored optimal designs for models that include only the main effects or only the main effects and two-factor interactions.

Turning to factorial designs for microarrays under the baseline parametrization, which is the main thrust of this paper, two key references are Yang and Speed (2002) and Glonek and Solomon (2004), (hereafter abbreviated GS). While Yang and Speed (2002) broadly discussed the design issues, GS reported illuminating computational results on admissible designs for the $2 \times 2$ factorial. Continuing with the baseline parametrization, we propose to consider general, possibly asymmetric, factorials and present analytical results. For comparative purposes, results under the orthogonal parametrization will also be occasionally indicated. The present endeavor is motivated by two reasons. First, as will be seen in Section 3, there are practical situations, even beyond the domain of microarray experiments, where the baseline parametrization is natural. Second, although this parametrization looks simpler than the orthogonal one, it renders the task of finding optimal or efficient designs somewhat more challenging due to lack of orthogonality. Presumably due to this reason, even in traditional factorial design literature, the optimal design problem under the baseline parametrization has received very little attention. It is hoped that our results would fill in this gap to some extent. We remark in this connection that because of the considerable difference in the definitions of the factorial effects under the baseline and orthogonal parametrizations, the significant body of work that has already been done on factorial microarray designs under the latter parametrization provides no clue to our derivation or results. The following example, concerning a study of leukemic mice, highlights this point; more details on some of the technical terms in the example are available in Sections 2 and 3.

EXAMPLE 1. GS describe a cDNA microarray experiment that compares two cell lines FI$\Delta$ and V449E at times 0 hours and 24 hours. The cell line V449E proliferates into leukemia while FI$\Delta$ is nonleukemic. Then there are two factors dictating the cell populations. The first factor, namely, the mutant, has two levels FI$\Delta$ and V449E of which FI$\Delta$, being nonleukemic, is taken as the baseline level. These levels are coded as 0 and 1 respectively. The second factor is time, again with two levels, 0 hours and 24 hours, and the first of these is taken as the baseline level. These two levels are also coded as 0 and 1 respectively. Thus, considering the two factors together, there are four treatment combinations, 00, 01, 10 and 11, representing the cell populations. The main effects of the two factors are important, but their interaction, that concerns the differential expression of genes for V449E and FI$\Delta$ at time 24 hours as contrasted with that at time 0 hours, is often of even greater interest. The experiment consists of a number of slides each comparing a pair of cell populations or, equivalently, treatment combinations. Suppose the available resources allow experimentation with six slides. Since the four treatment combinations can also be paired in six ways, namely, $(01, 00), (10, 00), (11, 00), (10, 01), (11, 01)$ and $(11, 10)$, the symmetric design, that compares each pair on one slide, seems very attractive, because it is a balanced incomplete block (BIB) design with excellent optimality properties under the orthogonal parametrization [Kiefer (1975)]. As a rival, consider the design that compares the pairs $(01, 00)$ and $(10, 00)$ each on two slides, and the pairs $(11, 01)$ and $(11, 10)$ each on one slide. As noted in GS, under the baseline parametrization, the symmetric design estimates the two main effects and the interaction with respective variances $\frac{1}{2}\sigma^2, \frac{1}{2}\sigma^2$ and $\sigma^2$, whereas the corresponding variances for the rival design are $\frac{5}{12}\sigma^2, \frac{5}{12}\sigma^2$ and $\frac{3}{4}\sigma^2$, where $\sigma^2$ is the common variance of the observations. Thus, the rival design outperforms the symmetric one not only in overall terms but also individually for each factorial effect, in the sense of entailing uniformly smaller variance, that is, more precise estimation.

The above example is revealing. It shows that even for the seemingly simple $2 \times 2$ factorial, optimal designs under the orthogonal parametrization are by no means guaranteed to perform well when one works with the baseline parametrization where entirely new designs may turn out to be desirable. This opens up new challenges which become even more complex for general factorials.

The paper is organized as follows. The next section gives an outline of the experimental setup. In Section 3 we revisit the $2 \times 2$ factorial, considered previously by GS, and obtain analytical results which supplement and strengthen their computational findings, in addition to preparing the ground for the subsequent development. Taking cognizance of the facts that in many applications there may be more than two factors dictating the cell populations and that, even with two factors, one or both of them may appear at more than two levels, general factorials are considered from Section 4 onward. Since cDNA microarray experiments are still quite expensive, there is a premium on optimal designs that are relatively small in size. From this viewpoint, in Section 4, we first consider the saturated, that is, most economic, case and present a collection of optimal designs in a strong sense. Apart from facilitating a choice under resource constraints, this leads to an approach for finding optimal or efficient designs in the practically more important nearly saturated cases that are also studied at length in Section 4. In Section 5 we extend the main ideas of Section 4 to the situation where the underlying model includes dye-coloring effects. The findings of this section rigorously justify, for such a model, a recommendation by Yang and Speed (2002) on dye-swapping. Several other design issues, including open problems, are discussed in Section 6. Technical details, including proofs, appear in a supplementary material file posted at the journal website [Banerjee and Mukerjee (2008)]. The technical tools include use of approximate design theory, Kronecker representation and unimodularity.

**2. Experimental setup.** We refer to Nguyen et al. (2002) and Amaratunga and Cabrera (2004) for detailed accounts of the experimental setup. In cDNA microarrays, each slide compares two cell populations on the basis of mRNA samples separately labeled with fluorescent dyes, usually red and green. This is done for a number of slides and different slides may compare different pairs of cell populations. After competitive hybridization, the ratio of the red and green fluorescence intensities is measured at each spot on each slide. Any such ratio represents the relative abundance of the gene in the two cell populations compared on the corresponding slide. The intensity ratios are usually adjusted for background noise and then normalized with the objective of removing systematic biases.

We consider linear models for the log intensities and, hence, the log intensity ratios. The modeling as well as the corresponding optimal design problem refers to a single gene—it is intended that the same design applies simultaneously to all genes on the array. The log intensity ratios for a gene, arising from different slides, are supposed to be homoscedastic and uncorrelated; a discussion on this, in the light of biological variability, follows in Section 6.

The above experimental setup is structurally similar to classical paired comparison experiments; see Kerr and Churchill (2001a). The cell populations under comparison are the same as treatments (or treatment combinations when they are dictated by several factors as in Example 1), while each slide is equivalent to a block of size two. However, the stringency on the number of slides as well as the baseline parametrization adopted here open up new design problems.

**3. The $2 \times 2$ factorial.**

3.1. *The baseline parametrization.* Suppose two factors $F_1$ and $F_2$, each at levels 0 and 1, dictate the cell populations, which correspond to the treatment combinations $00, 01, 10$ and $11$. Let $\tau_{00}, \tau_{01}, \tau_{10}$ and $\tau_{11}$ denote the expected log intensities, that is, the effects, of these treatment combinations. We focus on the situation where, as in Example 1, there is a null state or baseline level, say, 0, of each factor. Then $\theta_{00} = \tau_{00}$ stands for the baseline effect. We consider the baseline parametrization [cf. Yang and Speed (2002); GS] according to which the main effects of $F_1$ and $F_2$ are given respectively by

(1) $$\theta_{10} = \tau_{10} - \tau_{00} \quad \text{and} \quad \theta_{01} = \tau_{01} - \tau_{00},$$

while the interaction effect $F_1F_2$ is given by

$$\theta_{11} = \tau_{11} - \tau_{10} - \tau_{01} + \tau_{00}. \tag{2}$$

The counterparts of $\theta_{10}, \theta_{01}$ and $\theta_{11}$ under the more common orthogonal parametrization are defined respectively as

$$\theta_{10}^* = \tfrac{1}{2}(\tau_{11} + \tau_{10} - \tau_{01} - \tau_{00}), \qquad \theta_{01}^* = \tfrac{1}{2}(\tau_{11} - \tau_{10} + \tau_{01} - \tau_{00}),$$

$$\theta_{11}^* = \tfrac{1}{2}(\tau_{11} - \tau_{10} - \tau_{01} + \tau_{00}). \tag{3}$$

Observe that the definitions of the main effects under the two parametrizations are entirely different. While $\theta_{11}$ is proportional to $\theta_{11}^*$, this equivalence for the two-factor interaction also disappears for factorials involving three or more factors; see, e.g., (8) below.

Kerr (2006) nicely summarized the situations under which the two parametrizations mentioned above are appropriate. The baseline parametrization is natural if there is a clear null state or baseline level of each factor. As noted above, this happens in Example 1. Similarly, in a toxicological study with binary factors, each representing the presence or absence of a particular toxin, the state of absence can be regarded as a natural baseline level of each factor [Kerr (2006)]. On the other hand, if at least one factor, like gender, lacks a natural baseline level, then the baseline parametrization is inappropriate because this will arbitrarily single out one level of such a factor. In situations of this kind, it is advisable to use the orthogonal parametrization.

Indeed, the null state or baseline level of a factor can be interpreted in a broad sense. It need not strictly mean the zero level on some scale, but may as well refer to a standard or control level like the one currently used in practice. For example, in an agricultural experiment to investigate possible improvement in productivity by changing the doses of several fertilizers, the currently used doses of the fertilizers may represent the control levels. Similarly, in an industrial experiment on possible quality improvement via a change in the settings of several machines used at different stages of the production process, the current settings of the machines may reasonably constitute the control levels. In general, if each factor has such a control or baseline level along with one or more test levels, then the baseline parametrization is appropriate and, hence, the present results should be useful. The possible areas of application extend well beyond microarrays and pertain, notably, to agricultural and industrial experiments as hinted above. We add in this connection that although not much work has so far been reported on optimal factorial designs under the baseline parametrization, there is already a rich literature on the corresponding problem for varietal designs. An excellent review of this development on treatment-control designs is available in Majumdar (1996).

3.2. *Design criteria.* Following GS, in this section we assume the absence of systematic biases including dye-color bias because one of our objectives here is to obtain analytical results in their setup. With four treatment combinations $00, 01, 10$ and $11$, as in Example 1, there are six possibilities for any slide, namely, $(01, 00), (10, 00), (11, 00), (10, 01), (11, 01)$ and $(11, 10)$, where the members of each pair represent the treatment combinations that can be compared on the slide. Within each pair, one member gets red dye-coloring and the other green dye-coloring, but the distinction is immaterial in the absence of dye-color bias. Suppose the total number of slides used in the experiment is fixed at $N$. Then the design problem involves deciding on the respective frequencies $f_1, \ldots, f_6$ with which the slides of the six kinds as listed above should appear in the experiment, so as to entail optimal inference in a reasonable sense. Here $f_1, \ldots, f_6$ are nonnegative integers satisfying $f_1 + \cdots + f_6 = N$. We consider only those designs that keep $\theta_{01}, \theta_{10}$ and $\theta_{11}$ estimable. Let $V_{01}, V_{10}$ and $V_{11}$ denote the variances of the best linear unbiased estimators (BLUEs) of $\theta_{01}, \theta_{10}$ and $\theta_{11}$, respectively. A good design should aim at keeping these three variances small. Recognizing that commonly no single design will minimize all the three simultaneously, GS considered admissible deigns. A design $d_0$ is *admissible* if there is no other design $d_1$ such that each of $V_{01}, V_{10}$ and $V_{11}$ under $d_1$ is less than or equal to that under $d_0$, at least one of these inequalities being strict. By complete enumeration, GS tabulated admissible designs for even $N$ in the range $6 \leq N \leq 18$.

The notion of admissibility is intimately linked with that of weighted optimality. In most applications, one wishes to give equal weight to the two main effect parameters. Also, as GS noted, the interaction parameter can be of greater interest in microarrays than the main effect parameters. From this perspective, we consider designs that minimize $V_{01} + V_{10} + wV_{11}$, where $w$ is a positive weight, with particular interest in case $w > 1$. Such a design, called $w$-*optimal* for simplicity, is evidently admissible. Indeed, even for moderate $N$, admissible designs may be too numerous, and consideration of $w$-optimality helps in narrowing down the choice.

3.3. *Results via approximate design theory and their implications.* The fact that the frequencies $f_1, \ldots, f_6$ are integer-valued complicates the task of finding $w$-optimal designs because tools from calculus cannot be employed. This is particularly so because the objective function $V_{01} + V_{10} + wV_{11}$ depends on these frequencies in a complex manner. Considerable simplicity is achieved if for the moment we treat the relative frequencies $\pi_i = f_i/N$ as continuous variables over the range $\pi_i \geq 0$ for each $i$ and $\sum \pi_i = 1$. Any such $\pi = (\pi_1, \ldots, \pi_6)$ is called a design measure. This approach amounts to invoking the approximate design theory [see, e.g., Silvey (1980)] which enables one to use calculus techniques to get the following result.

RESULT 1. (a) For $w > 0$, let

(4) $$\xi = \tfrac{1}{4}\{(w^2 + 2w)^{1/2} - w\}.$$

Then the design measure

(5) $$\pi_0 = (\tfrac{1}{2} - \xi, \tfrac{1}{2} - \xi, 0, 0, \xi, \xi)$$

is $w$-optimal, whenever $w \geq \tfrac{2}{3}$.

(b) The design measure $\tilde{\pi} = (\tfrac{1}{4}, \tfrac{1}{4}, 0, 0, \tfrac{1}{4}, \tfrac{1}{4})$ minimizes $V_{11}$ and is admissible.

(c) The design measure $(\tfrac{1}{2} - \xi, \tfrac{1}{2} - \xi, 0, 0, \xi, \xi)$ is admissible whenever $\tfrac{1}{6} \leq \xi \leq \tfrac{1}{4}$.

Incidentally, Bueno Filho, Gilmour and Rosa (2006) and Grossmann and Schwabe (2008) also employed the approximate theory in the study of optimal microarray designs. But their settings and criteria and, hence, final results are different from ours. We now discuss the implication of Result 1 on (exact) designs that take cognizance of the fact that $f_1, \ldots, f_6$ are actually integers. Any such design may be represented by the vector $f = (f_1, \ldots, f_6)$. Since $\pi_i = f_i/N$ for each $i$, the following conclusions, pertaining to *even N*, are evident from Result 1.

(i) If $w \geq \tfrac{2}{3}$ and $\phi = N\xi$ is an integer, where $\xi$ is given by (4), then the design

(6) $$f = (\tfrac{1}{2}N - \phi, \tfrac{1}{2}N - \phi, 0, 0, \phi, \phi)$$

is $w$-optimal.

(ii) If $N$ is a multiple of 4, then the design $(\tfrac{1}{4}N, \tfrac{1}{4}N, 0, 0, \tfrac{1}{4}N, \tfrac{1}{4}N)$ minimizes $V_{11}$.

(iii) Any design of the form (6) is admissible whenever $\tfrac{1}{6}N \leq \phi \leq \tfrac{1}{4}N$.

The points (i)–(iii) noted above cater to the need, in our context, of finding good designs with emphasis on the interaction parameter. For $N \leq 18$, they provide analytical justification for quite a few findings of GS, such as the admissibility of the rival design in Example 1. In addition, they facilitate the study of good designs for $N \geq 20$, which is beyond the range considered by GS and may pose difficulties in complete enumeration. For instance, if $N = 20$, then they show that the designs $(6, 6, 0, 0, 4, 4)$ and $(5, 5, 0, 0, 5, 5)$ are admissible, and that the latter design minimizes $V_{11}$.

Given $w$ ($\geq \tfrac{2}{3}$), even if $\phi = N\xi$ in (i) above is not an integer, one may simply round it off to the nearest integer to get a highly efficient design. As an illustration, let $w = 2$. By (4), then $\xi = 0.207107$. For $N = 22$, rounding $N\xi$ off to the nearest integer, namely, 5, we can follow (6) to consider the design $(6, 6, 0, 0, 5, 5)$, which has efficiency 99.44% as a comparison with the $w$-optimal design measure in (5) reveals. Continuing with $w = 2$, for every even $N$ in the range $6 \leq N \leq 30$, one can similarly obtain designs with over 97%, and often over 99%, efficiency. These efficiencies are actually lower bounds, as they are relative to an optimal deign measure which is unattainable in the exact setup. Hence, we conjecture that all these designs are actually $w$-optimal, with $w = 2$, for the respective $N$. Using (iii) above, one can also verify that these designs are all admissible.

It is of interest to compare Result 1 with its counterpart arising under the orthogonal parametrization (3). To that effect, we note that the following hold under (3):

(a) The design measure $\pi_0^{\text{orth}} = (\alpha, \alpha, \frac{1}{2} - 2\alpha, \frac{1}{2} - 2\alpha, \alpha, \alpha)$, where $\alpha = \frac{1}{2} w^{1/2}/(2 + w^{1/2})$, is $w$-optimal for $0 < w < 4$.

(b) The design measure $\tilde{\pi} = (\frac{1}{4}, \frac{1}{4}, 0, 0, \frac{1}{4}, \frac{1}{4})$ is $w$-optimal for $w \geq 4$.

(c) The design measure $(\alpha, \alpha, \frac{1}{2} - 2\alpha, \frac{1}{2} - 2\alpha, \alpha, \alpha)$, with $\alpha$ as in (a), is admissible for $0 < \alpha \leq \frac{1}{4}$.

The proofs of these are similar to but simpler than that of Result 1. From (a) and (b) above, simple rounding off again yields highly efficient exact designs under the orthogonal parametrization. For $w < 4$, unlike $\pi_0$ in Result 1, the measure $\pi_0^{\text{orth}}$ in (a) assigns positive masses to all the six possible slides. In fact, for $w = 1, \pi_0^{\text{orth}}$ assigns uniform mass everywhere and, hence, entails a BIB design. On the other hand, for $w \geq 4$, the optimal design measures under the two parametrizations are quite close to each other.

The fact that $\pi$ has only six elements, because of only six possibilities for any slide, helped the study of optimal design measures and, hence, that of optimal or efficient designs in this section. In microarray experiments for general factorials considered from Section 4 onward, the number of possibilities for any slide increases dramatically and, as a result, the optimal design measures are analytically intractable. For this reason, hereafter we directly investigate exact designs.

## 4. General factorials.

4.1. *Preliminaries.* In many applications of cDNA microarrays there may be more than two factors dictating the cell populations and, even if there are only two factors, one or both of them may appear at more than two levels. For instance, if in Example 1 the two cell lines are compared at time 12 hours, in addition to 0 hours and 24 hours, then we have to consider a $2 \times 3$ factorial, with the second factor, time, now appearing at three levels. Similar examples abound and underscore the practical need to explore the optimal designing of cDNA microarray experiments with reference to general factorials.

From this perspective, consider an $s_1 \times \cdots \times s_n$ factorial that involves $n \, (\geq 2)$ factors $F_1, \ldots, F_n$ dictating the cell populations, with $F_j$ appearing at levels $0, 1, \ldots, s_j - 1$. Then there are $v = \prod s_j$ cell populations which correspond to the treatment combinations $i_1 \ldots i_n$ $(0 \leq i_j \leq s_j - 1, 1 \leq j \leq n)$. Let $\tau_{i_1 \ldots i_n}$ be the expected log intensity, that is, the effect, of the treatment combination $i_1 \ldots i_n$. As before, the baseline level of each factor is denoted by 0. Hence, $\theta_{00 \ldots 0} = \tau_{00 \ldots 0}$ stands for the baseline effect. Also, as an obvious extension of the baseline parametrization given by (1) and (2), a main effect, say, that of $F_1$, is represented by the $s_1 - 1$ parameters

$$\theta_{i_1 0 \ldots 0} = \tau_{i_1 0 \ldots 0} - \tau_{00 \ldots 0} \qquad (1 \leq i_1 \leq s_1 - 1), \tag{7}$$

whereas a two-factor interaction, say, $F_1 F_2$, is represented by the $(s_1 - 1)(s_2 - 1)$ parameters

$$\theta_{i_1 i_2 0 \ldots 0} = \tau_{i_1 i_2 0 \ldots 0} - \tau_{i_1 00 \ldots 0} - \tau_{0 i_2 0 \ldots 0} + \tau_{000 \ldots 0} \tag{8}$$

$$(1 \leq i_1 \leq s_1 - 1, 1 \leq i_2 \leq s_2 - 1).$$

Similarly, we can define $\theta_{i_1 \ldots i_n}$ for every $i_1 \ldots i_n \neq 0 \ldots 0$ $(0 \leq i_j \leq s_j - 1, 1 \leq j \leq n)$ so that any such $\theta_{i_1 \ldots i_n}$ represents a factorial effect as determined by its nonzero subscripts. Thus, any $\theta_{i_1 \ldots i_n}$ with $u$ nonzero subscripts represents a factorial effect involving $u$ factors. Hereafter, often the $v - 1$ parameters $\theta_{i_1 \ldots i_n}$ $(i_1 \ldots i_n \neq 0 \ldots 0)$ are collectively referred to as the $\theta$s for ease in presentation.

Note that (7) is reminiscent of the canonical parametrization in Wit, Nobile and Khanin (2005), Subsection 2.1, for varietal designs. Throughout this section we continue to assume the absence of systematic biases including dye-color bias and write $\sigma^2$ for the variance of any observed log intensity ratio.

4.2. *Optimal saturated designs.* All main and interaction effects, as represented by the $\theta$s, are of potential interest at least for a relatively small number of factors. Hence, at this stage we consider optimal designs for the estimation of all these $v - 1$ parameters. Clearly, then the number of slides, $N$, in the experiment must satisfy $N \geq v - 1$. We first consider the saturated case $N = v - 1$ and obtain a collection of optimal designs. In addition to facilitating a choice under resource constraints, this paves the way for the development of an approach for finding optimal or efficient designs in the practically more important nearly saturated cases that are taken up in the next subsection.

RESULT 2. *Let $N = v - 1$ and consider a design that keeps all the $\theta$s estimable. Then for any $\theta_{i_1 \ldots i_n}$, which represents a factorial effect involving $u$ factors,*

$$\mathrm{Var}(\hat{\theta}_{i_1 \ldots i_n}) \geq \sigma^2 2^{u-1}, \tag{9}$$

where $\hat{\theta}_{i_1...i_n}$ is the BLUE of $\theta_{i_1...i_n}$.

Result 3 below shows that the same design can attain the lower bound in (9) simultaneously for all the $\theta$s. Such a design is then optimal not only in overall terms but also *individually* for every parameter representing a main or interaction effect. In what follows, a slide which compares treatment combinations $i_1\ldots i_n$ and $j_1\ldots j_n$, respectively with red and green dye-coloring, is denoted by the ordered pair $(i_1\ldots i_n, j_1\ldots j_n)$. A design is represented by a collection of such pairs. Note that the ordering within any pair is immaterial at this stage for inferential purposes, as we are now assuming the absence of dye-color bias. For any $i_1\ldots i_n \ne 0\ldots 0$, let $\rho(i_1\ldots i_n)$ be obtained replacing the first nonzero entry of $i_1\ldots i_n$ by 0 and leaving the other entries unchanged. For instance, with a $2\times 2\times 3$ factorial, $\rho(012)=002, \rho(111)=011$ etc.

RESULT 3. Let $N = v - 1$. Then the design

$$d_0 = \{(i_1\ldots i_n, \rho(i_1\ldots i_n)) : 0 \le i_j \le s_j - 1, 1 \le j \le n, i_1\ldots i_n \ne 0\ldots 0\}$$

leads to the attainment of the lower bound in (9) simultaneously for all the $\theta$s.

REMARK 1. For $n=2$, one can check that Result 3 remains valid if for every $i_1 \ge 1, i_2 \ge 1, \rho(i_1 i_2)$ is allowed to be either $0i_2$ or $i_1 0$, instead of being fixed at $0i_2$ as stipulated above. Because of the two possibilities for any such $\rho(i_1 i_2)$, one gets a collection of $2^{(s_1-1)(s_2-1)}$ optimal designs.

REMARK 2. Examples can be given to show that, for $n \ge 3$, Result 3 does not remain valid if, in the spirit of Remark 1, each $\rho(i_1\ldots i_n)$ is obtained simply by replacing an arbitrary, rather than the first, nonzero entry of $i_1\ldots i_n$ by 0. However, even then, Result 3 leads to a collection of optimal designs via factor permutation. To illustrate this point, observe that Result 3 yields the design

$$d_0 = \{(001, 000), (002, 000), (010, 000), (011, 001), (012, 002), (100, 000),$$
$$(101, 001), (102, 002), (110, 010), (111, 011), (112, 012)\}$$

for a $2 \times 2 \times 3$ factorial, and the design

$$\{(001, 000), (010, 000), (011, 001), (100, 000), (101, 001), (110, 010),$$
$$(111, 011), (200, 000), (201, 001), (210, 010), (211, 011)\}$$

for a $3 \times 2 \times 2$ factorial. Permuting the factors in the latter, one readily gets another design for the $2 \times 2 \times 3$ factorial which, like $d_0$, is optimal in the sense of Result 3. In the same manner, Result 3 can be easily applied to all possible factor orderings to yield a collection of optimal designs.

4.3. *Nearly saturated designs.* The optimal designs Section 4.2 are saturated and, hence, do not yield an internal estimator of $\sigma^2$ which is important for testing of hypotheses. This difficulty can persist even if the same clone is replicated $r$ ($>1$) times on each slide. Then, for the purpose of estimating the $\theta$s, the means of the $r$ log intensity ratios arising from the slides play the role of the individual ratios considered so far, but an attempt to estimate $\sigma^2$ on the basis of the within slide variation can be vitiated by unknown correlation among the ratios arising from the same slide [Yang and Speed (2002) and Churchill (2002)].

In view of the above, as a feasible yet economic approach to getting degrees of freedom for the estimation of $\sigma^2$, one may like to have a little more than $v-1$ slides and thus consider nearly saturated designs. Unlike in the saturated case, typically for $N > v - 1$, no single design can estimate all the $\theta$s with the minimum variance. We, therefore, consider the $w$-optimality criterion as applicable to general factorials. Analytical derivation of optimal designs, via either combinatorial techniques or approximate design theory, still remains difficult for $N > v - 1$.

The results in the last subsection, however, readily yield a heuristic approach which, as computations indicate, leads to highly efficient, if not optimal, designs. Based on the intuitive expectation that, for $N$ close to $v-1$, a design obtained via augmentation of an optimal saturated design should behave well, we propose the following steps:

(I) Given $s_1, \ldots, s_n$, list all optimal saturated designs given by Result 3 and Remark 1 or 2.

(II) Given $N$, augment each design in (I) in all possible ways to generate designs with $N$ slides.

(III) From the augmented designs in (II), select one as per the chosen optimality criterion, also taking care of resource constraints, if any.

For $N$ close to $v - 1$, computationally the above steps are far easier to implement than a complete enumeration of all designs. The least favorable cases for this approach are those where the saturated designs involve a rather small number of slides, and hence, even a slightly larger $N$ can potentially have a significant impact. From this viewpoint, the approach is now evaluated for $2 \times 2$ and $2 \times 3$ factorials which represent the two smallest saturated cases. We consider $w$-optimal designs that, for an $s_1 \times s_2$ factorial, aim at minimizing

$$\sum_{i_1=1}^{s_1-1} \text{Var}(\hat{\theta}_{i_1 0}) + \sum_{i_2=1}^{s_2-1} \text{Var}(\hat{\theta}_{0 i_2}) + w \sum_{i_1=1}^{s_1-1} \sum_{i_2=1}^{s_2-1} \text{Var}(\hat{\theta}_{i_1 i_2}).$$

Tables 1 and 2 show $w$-optimal designs, as obtained by complete enumeration of all designs, for $2 \times 2$ and $2 \times 3$ factorials, with $w = 1, 2, 3$, and $N = v -$ $1 + j (j = 1, 2, 3)$. While these optimal designs can be nonunique, only one such design is reported in each case to save space.

TABLE 1
*w-optimal designs for the $2 \times 2$ factorial*

| $N$ | $w$ | $w$-optimal design |
|---|---|---|
| 4 | 1, 2, 3 | $\{(01,00),(10,00),(11,01),(11,10)\}$ |
| 5 | 1, 2, 3 | $\{(01,00),(10,00),(10,00),(11,01),(11,10)\}$ |
| 6 | 1, 2, 3 | $\{(01,00),(01,00),(10,00),(10,00),(11,01),(11,10)\}$ |

Every design in Table 1 or 2 contains the optimal saturated design $d_0$ of Result 3 as a subdesign, thus showing that the heuristic approach, based on augmentation, indeed yields a $w$-optimal design in each of these cases. For the $2 \times 2$ factorial, we have also checked that all admissible designs for $N = 4, 5$ and 6 are augmentations of one of the two optimal saturated designs, $\{(01,00),(10,00),(11,01)\}$ and $\{(01,00),(10,00),(11,10)\}$, arising from Remark 1.

For an $s_1 \times s_2$ factorial, let $\bar{d}$ be the design obtained as the union of the $2^{(s_1-1)(s_2-1)}$ optimal saturated designs given by Remark 1, that is, $\bar{d}$ consists of the $v - 1 + (s_1 - 1)(s_2 - 1)$ slides $(i_1 i_2, 0 i_2), 1 \leq i_1 \leq s_1 - 1, 0 \leq i_2 \leq s_2 - 1$, and $(i_1 i_2, i_1 0), 0 \leq i_1 \leq s_1 - 1, 1 \leq i_2 \leq s_2 - 1$. Let $\Omega$ be the class of all designs that consist of $N$ slides from $\bar{d}$, where $v - 1 < N \leq v - 1 + (s_1 - 1)(s_2 - 1)$. Tables 1, 2 and partial enumeration in several other cases lead us to the following conjecture.

CONJECTURE. (a) If $N = v - 1 + (s_1 - 1)(s_2 - 1)$, then the design $\bar{d}$ is $w$-optimal for any $w \geq 1$.

(b) If $v - 1 < N < v - 1 + (s_1 - 1)(s_2 - 1)$, then for any $w \geq 1$, a $w$-optimal design in $\Omega$ is also $w$-optimal in the class of all designs.

The case $N = 4$ in Table 1 and the cases $N = 6, 7$ in Table 2 pertain to the Conjecture and show its truth, with $w = 1, 2, 3$, for $2 \times 2$ and $2 \times 3$ factorials. Furthermore, using approximate design theory, the efficiency of the design $\bar{d}$ in (a) is seen to be at least 91.88%, 94.16% and 95.57% for the $2 \times 4$ factorial, and at least 94.26%, 96.16% and 97.04% for the $3 \times 3$ factorial, under $w = 1, 2$ and 3 respectively. Indeed, if this Conjecture is true in general, then part (b) would considerably reduce the search for an optimal design, while part (a) would give a compact result.

TABLE 2
*w-optimal designs for the $2 \times 3$ factorial*

| $N$ | $w$ | $w$-optimal design |
|---|---|---|
| 6 | 1, 2, 3 | $\{(01,00),(02,00),(10,00),(11,01),(12,02),(11,10)\}$ |
| 7 | 1, 2, 3 | $\{(01,00),(02,00),(10,00),(11,01),(12,02),(11,10),(12,10)\}$ |
| 8 | 1 | $\{(01,00),(02,00),(10,00),(11,01),(12,02),(11,10),(12,10),(02,01)\}$ |
| 8 | 2, 3 | $\{(01,00),(02,00),(10,00),(10,00),(11,01),(12,02),(11,10),(12,10)\}$ |

For general factorials, one can define the orthogonal parametrization via a straightforward extension of (3); see, for example, Gupta and Mukerjee [(1989), Chapter 2]. Under this parametrization, BIB designs are optimal in a very strong sense [Kiefer (1975)] and extended group divisible (EGD) designs are known to be admissible [Gupta and Mukerjee (1989), Chapters 3 and 8]. While Example 1 demonstrates that a BIB design can become inadmissible under the baseline parametrization, we now show that the same can happen with EGD designs. Note that in the context of microarrays, an EGD design is one where the number of slides comparing any two treatment combinations $i_1 \ldots i_n$ and $j_1 \ldots j_n$ depends only on the equality or otherwise of $i_u$ and $j_u, 1 \le u \le n$. Thus, for the $2 \times 3$ factorial and with $N = 6$ slides, there is a unique EGD design $\{(11,00),(12,00),(10,01),(12,01),(10,02),(11,02)\}$ that allows the estimability of all treatment contrasts. Under the baseline parametrization, this design becomes inadmissible because it estimates each of the $\theta$s with uniformly larger variance than the $w$-optimal design shown in Table 2 for $N = 6$.

**5. Results under effects due to dye-coloring.** We now extend the main ideas of Section 4 to the situation where the underlying model includes effects due to dye-coloring. For $0 \le i_j \le s_j - 1, 1 \le j \le n$, let $\beta^{(1)}_{i_1 \ldots i_n}$ and $\beta^{(2)}_{i_1 \ldots i_n}$ be the expected log intensities for the treatment combination $i_1 \ldots i_n$ under red and green dye-coloring respectively. Then $\tau_{i_1 \ldots i_n} = \frac{1}{2}\{\beta^{(1)}_{i_1 \ldots i_n} + \beta^{(2)}_{i_1 \ldots i_n}\}$ represents the overall effect of $i_1 \ldots i_n$, whereas $\lambda_{i_1 \ldots i_n} = \frac{1}{2}\{\beta^{(1)}_{i_1 \ldots i_n} - \beta^{(2)}_{i_1 \ldots i_n}\}$ accounts for the effect of dye-coloring on $i_1 \ldots i_n$. For any slide $(i_1 \ldots i_n, j_1 \ldots j_n)$, which compares treatment combinations $i_1 \ldots i_n$ and $j_1 \ldots j_n$ respectively with red and green dye-coloring, the expected log intensity ratio is now given by

$$(10) \qquad \beta^{(1)}_{i_1 \ldots i_n} - \beta^{(2)}_{j_1 \ldots j_n} = \tau_{i_1 \ldots i_n} - \tau_{j_1 \ldots j_n} + \lambda_{i_1 \ldots i_n} + \lambda_{j_1 \ldots j_n}.$$

The parameters of interest continue to be the $\theta$s, representing the main and interaction effects and defined with reference to the $\tau$s as in Section 4.1. The $\lambda$s are, on the other hand, nuisance parameters to us. Unlike in the previous sections, where we took the $\lambda$s as zeros, now these are kept perfectly general. Hence, as (10) indicates, the ordering within the slides is no longer inconsequential. A reduced but more restrictive version of the model (10) will be considered briefly in Section 6.

In the presence of dye-coloring effects, several authors, notably Yang and Speed (2002), advocated the use of dye-swapped experiments. It is not hard to see that, under (10), any estimable contrast of the $\tau$s is estimated orthogonally to the $\lambda$s in such an experiment. Let $d_0$ be any optimal design arising from Result 3, Remark 1 or Remark 2. The dye-swapped version of $d_0$, denoted by $d_0^{\text{swap}}$, is a design that includes both the slides $(i_1 \ldots i_n, j_1 \ldots j_n)$ and $(j_1 \ldots j_n, i_1 \ldots i_n)$ for every slide $(i_1 \ldots i_n, j_1 \ldots j_n)$ in $d_0$. Given the optimality of $d_0$ in the absence of dye-coloring effects, one may be inclined to recommend the use of $d_0^{\text{swap}}$ in the present setup. However, in order to justify this rigorously, the following questions need to be answered:

(a) There are $2(v-1)$ slides in $d_0^{\text{swap}}$. Are at least $2(v-1)$ slides required to estimate all the $\theta$s under (10), even when possibly nonorthogonal (to the $\lambda$s) estimation is allowed?

(b) Under (10), will $d_0^{\text{swap}}$ be optimal, in the sense of Result 3, among all designs that involve $2(v-1)$ slides and keep the $\theta$s estimable?

The possibility of nonorthogonal estimation complicates (a). Similarly, (b) needs careful attention because orthogonality alone does not guarantee optimality with $2(v-1)$ slides.

Satisfyingly, the answers to both (a) and (b) are in the affirmative. The following results confirm this and, hence, vindicate the proposal of Yang and Speed (2002) about dye-swapping. As before, the total number of slides is denoted by $N$. Also, we continue to assume that the log intensity ratios arising from different slides are uncorrelated and homoscedastic with common variance $\sigma^2$.

RESULT 4. *Under the model* (10), *at least* $2(v-1)$ *slides are required to keep all the $\theta$s estimable.*

RESULT 5. *Let* $N = 2(v-1)$ *and consider a design that keeps all the $\theta$s estimable under* (10). *Then for any* $\theta_{i_1 \ldots i_n}$, *which represents a factorial effect involving $u$ factors,*

$$\text{Var}(\hat{\theta}_{i_1 \ldots i_n}) \geq \sigma^2 2^{u-2}, \tag{11}$$

*where* $\hat{\theta}_{i_1 \ldots i_n}$ *is the BLUE of* $\theta_{i_1 \ldots i_n}$ *under* (10).

RESULT 6. *Let* $N = 2(v-1)$ *and* $d_0^{\text{swap}}$ *be a design defined as above. Then, under* (10), $d_0^{\text{swap}}$ *leads to the attainment of the lower bound in* (11) *simultaneously for all the $\theta$s.*

These results show the optimality of $d_0^{\text{swap}}$ in a strong sense. Although Result 5 resembles Result 2, its proof involves much extra work. Following Remarks 1 and 2, there is considerable flexibility in the choice of $d_0$ and hence that of $d_0^{\text{swap}}$. In addition to being helpful under resource constraints, this facilitates the task of finding highly efficient designs when one intends to use a little more than $2(v-1)$ slides so as to gain degrees of freedom for the estimation of $\sigma^2$. For this purpose, the same heuristic approach as in Section 4.3 can be followed with the only change that now in step (I), all possibilities for $d_0^{\text{swap}}$, corresponding to $d_0$ arising from Result 3 and Remark 1 or 2, have to be considered. As an illustration, consider the $2 \times 2$ factorial and let $N = 8$. Then, under the criterion of $w$-optimality ($w = 1, 2$ or 3), the above approach yields the design

$$\{(01,00),(10,00),(11,01),(00,01),(00,10),(01,11),(11,10),(10,11)\},$$

which is an augmentation of $d_0^{\text{swap}}$ (consisting of the first six slides) and a dye-swapped design by itself. A complete enumeration shows that this design is, indeed, $w$-optimal among all designs with $N = 8$ slides, for $w = 1, 2, 3$ and under the model (10).

## 6. Concluding remarks.

6.1. *Robustness considerations.* The results in this paper were obtained under the assumption that the log intensity ratios for a gene, arising from different slides, are homoscedastic and uncorrelated. A discussion on this assumption is warranted. In cDNA microarray experiments, the measurement error is typically swamped in biological variability. From a practical viewpoint, it is therefore appropriate to attribute the variance of an observed log intensity ratio arising from a slide $(i_1 \ldots i_n, j_1 \ldots j_n)$ to components, say, $\gamma_{i_1 \ldots i_n}^2$ and $\gamma_{j_1 \ldots j_n}^2$, representing the biological variability within the cell populations given by $i_1 \ldots i_n$ and $j_1 \ldots j_n$, in addition to a component, say, $\delta^2$, due to the measurement error. Thus, this variance equals $\gamma_{i_1 \ldots i_n}^2 + \gamma_{j_1 \ldots j_n}^2 + \delta^2$. If the variance components $\gamma_{i_1 \ldots i_n}^2$ are supposed to be equal for all cell populations [cf. Kerr (2003) and Altman and Hua (2006), among others] with common value say, $\gamma^2$, then the log intensity ratios arising from different slides are homoscedastic with common variance $\sigma^2 = 2\gamma^2 + \delta^2$. Furthermore, these ratios can be safely assumed to be uncorrelated if the replications for every treatment combination are biological (i.e., the same subject does not appear in more than one slide). Thus, in this situation all our results go through with $\sigma^2 = 2\gamma^2 + \delta^2$.

If, however, the variance components $\gamma_{i_1 \ldots i_n}^2$ associated with the cell populations are not all equal, then the assumption of homoscedasticity no longer holds. In order to give a flavor of the robustness of our results to this possibility, we revisit Sections 3 and 4. For the $2 \times 2$ factorial in Section 3, writing $\tilde{\gamma}_{i_1 i_2}^2 = \gamma_{i_1 i_2}^2 / \delta^2$, three patterns are considered for $(\tilde{\gamma}_{00}^2, \tilde{\gamma}_{01}^2, \tilde{\gamma}_{10}^2, \tilde{\gamma}_{11}^2)$: (i) $(2, 2.5, 2.5, 3)$, (ii) $(2, 3, 4, 6)$ and (iii) $(6, 4, 3, 2)$. Under (i)–(iii), one can employ the approximate design theory to find the $w$-optimal design measures and, hence, obtain Table 3 showing the efficiencies of the design measure $\pi_0$ reported earlier in Result 1(a). It is satisfying to note that $\pi_0$, which is $w$-optimal under homoscedasticity, remains quite robust even to the appreciably heteroscedastic situations (ii) and (iii). The exact designs arising from $\pi_0$ are also seen to remain highly efficient under (i)–(iii). The findings are almost equally encouraging for the nearly saturated optimal exact designs shown in Tables 1 and 2. Under (i)–(iii) and for $w = 1, 2, 3$, the designs in Table 1 often remain $w$-optimal among all exact designs and, except in one case, always have efficiency over 97%. The exceptional case concerns the design for $N = 6$, which has efficiency 93.40% under (ii) when $w = 3$. For the $2 \times 3$ factorial, along the line of (i)–(iii), we considered the patterns $(2, 2.5, 2.5, 2.5, 3, 3)$, $(2, 3, 3, 4, 6, 6)$ and $(6, 4, 4, 3, 2, 2)$ for $(\tilde{\gamma}_{00}^2, \tilde{\gamma}_{01}^2, \tilde{\gamma}_{02}^2, \tilde{\gamma}_{10}^2, \tilde{\gamma}_{11}^2, \tilde{\gamma}_{12}^2)$. Under all these patterns and for $w = 1, 2, 3$, the designs in Table 2 often turn out to be $w$-optimal and always have efficiency over 98%.

TABLE 3
*Efficiencies of $\pi_0$ under heteroscedasticity*

| Situation | $w = 1$ | $w = 2$ | $w = 3$ |
|---|---|---|---|
| (i)   | 99.90% | 99.89% | 99.86% |
| (ii)  | 99.25% | 99.03% | 98.91% |
| (iii) | 99.43% | 99.15% | 99.00% |

The log intensity ratios from different slides, of course, get correlated when the same subject is allowed to appear in more than one slide. If we continue to assume the equality of the $\gamma_{i_1 \ldots i_n}^2$ and denote their common value by $\gamma^2$, then the correlation terms depend on the ratio $\gamma^2/\delta^2$, which is commonly unknown. As a result, the standard linear model based analysis and the associated optimal design theory will not work. On the other hand, if we pretend $\gamma^2/\delta^2$ to be known so as to allow the use of weighted least squares, empirical evidence suggests in favor of having only biological replications from the point of view of efficiency. To illustrate this point without making the presentation too involved, we consider the case of a $2 \times 2$ factorial design in $N = 4$ slides under the absence of dye color effects. We made an enumeration of all such designs that keep the main and interaction effects estimable and, for each design and every treatment combination, enumerated all possibilities for biological or technical replication (here technical replication means repeating the same subject on more than one slide). For instance, in the design $d^* = \{(01, 00), (10, 00), (11, 01), (11, 10)\}$, the two replications for any treatment combination can be biological or technical, thus leading to 16 possibilities arising from this design alone. The complete enumeration revealed that, if the ratio $\gamma^2/\delta^2$ is pretended to be known, then irrespective of the value of this ratio, the design $d^*$, with all replicates biological, is $w$-optimal whenever $w \geq \frac{2}{3}$. Earlier, in Table 1, the same design was reported to be $w$-optimal for $w = 1, 2$ and 3 in the homoscedastic and uncorrelated setup. Complete enumeration of this kind becomes unmanageable for more complex factorials, but partial enumeration in several other situations led to similar conclusions. This reinforces the findings in Kerr (2003) in a simpler setting and suggests that, in addition to making the log intensity ratios from different slides uncorrelated, use of only biological replicates can be advantageous from the perspective of design efficiency as well; see also Kendziorski et al. (2005) and the references therein for insightful practical results in a similar context. The point just noted makes sense if the cost of biological replication is negligible compared to the cost of the assay per slide, as has been tacitly supposed in this paper. While Bueno Filho, Gilmour and Rosa (2006) mention that the number of slides is typically the most important limiting factor in microarray experiments, a more detailed discussion in this regard is available in Kerr (2003), who also dwelt on the situation where this is not the case. If the cost of biological replication is a real issue, then the design problem becomes much more complex. Instead of fixing the number of slides, as done here, one should then proceed in the spirit of Kerr (2003) to formulate the problem in terms of a cost function that incorporates the cost of the assays (slides), as well as the cost of biological replication. Given such a cost constraint, the possibility of technical replication and associated correlation will also have to be accounted for. Since commonly this correlation is unknown, the optimal design problem will then concern some kind of likelihood based rather than linear model based analysis.

6.2. *Further open issues.* Even within the homoscedastic and uncorrelated setting, there are several open issues that deserve attention. One of these concerns analytical derivation of optimal designs for $N$ greater than $v - 1$ or $2(v - 1)$ in Sections 4 or 5 respectively. For instance, a proof of the Conjecture in Section 4.3 will be of interest. This can, however, be challenging, and pending a complete solution, our heuristic approach holds the promise of yielding designs that are at least highly efficient.

From a practical point of view, an important design problem is that of fractional replication. Compared to traditional factorials, a difficulty here is the lack of effect hierarchy [Wu and Hamada (2000), page 112]. Even in the two-factor case, the interaction can be of greater interest to biologists than the main effects. Hence, especially when the number of factors, $n$, is relatively small, it may be too drastic to ignore some interactions, as required in fractional replication. For large $n$, however, this can be a sensible option. The experience with factorial fractions under the orthogonal parametrization [see, e.g., Dey and Mukerjee (1999)] suggests that then, under specification from biologists about the pattern of negligible interactions, the present techniques should be useful.

A problem, akin to that of fractional replication, concerns the study of optimal deigns when the impact of possible dye-color bias can be modeled via a reduced version of (10). Note that the model (10) allows a very general form for the effect of dye-coloring and, hence, is applicable to a broad spectrum of situations. If in a specific application one has sufficient knowledge of the underlying process so as to entertain the risk of assuming that such effect is repeatable over slides, that is, additive to treatment effects, then in (10) one can replace $\lambda_{i_1 \ldots i_n} + \lambda_{j_1 \ldots j_n}$ by a single parameter $\eta$. In this case, it can be shown that at least $v$ slides are required to keep all the $\theta$s estimable. However, in contrast with Results 3 and 6, no single design with $v$ slides is optimal simultaneously for all these parameters. For this reduced model, it is known that any even design (i.e., a design where every treatment combination appears an even number of times) allows a dye-color assignment that ensures orthogonality to $\eta$ [Kerr and Churchill (2001a)]. For even $s_1$ and $s_2$, the design in Conjecture (a) of Section 4.3 is even and, hence, with appropriate dye-color assignment, it is again conjectured to be optimal under this model. For odd $s_1$ or $s_2$ too, the initial findings are optimistic. Thus, for the $2 \times 3$ factorial, Conjecture (a) yields the nearly orthogonal design $\{(00, 01), (02, 00), (10, 00), (01, 11), (11, 10), (12, 02), (10, 12)\}$, and a complete enumeration shows that it is, indeed, $w$-optimal among all designs with 7 slides, for $w = 1, 2, 3$ and under the reduced model.

In the present paper we studied optimal designs from the statistical consideration of efficiency. From this perspective, our designs often outperform more elementary ones that have gained popularity in applied work. For instance, in the setup of Sections 4 and 5, it is easy to check that the designs arising from Results 3 and 6 estimate the main effect parameters with the same variance and entail smaller variances for the interaction parameters, as compared to the commonly used reference design or the dye-swapped version thereof, respectively. These simpler designs may, however, have other practical benefits, including those dictated by manufacturer recommendations. Nevertheless, as noted earlier, our results allow considerable flexibility under resource constraints and should be useful to applied researchers concerned with practical issues in addition to efficiency considerations. Even in extreme situations where such practicalities preclude direct implementation of the proposed designs, the latter would help in benchmarking the designs actually used from the point of view of efficiency.

It is hoped that the present endeavor will generate further interest in the above directions.

**Acknowledgments.** We thank the referee, the associate editor and the editor for very constructive suggestions. This work was supported by the Center for Management and Development Studies, Indian Institute of Management Calcutta.

## SUPPLEMENTARY MATERIAL

**Optimal factorial designs for CDNA microarray experiments: Proofs** (doi: 10.1214/07-AOAS144SUPP; .pdf). Technical details, including proofs, appear in a supplementary material file posted at the journal website.

| | |
|---|---|
| Indian Institute of Management Ahmedabad | Indian Institute of Management Calcutta |
| Vastrapur | Joka |
| Ahmedabad 380 015 | Diamond Harbour Road |
| India | Kolkata 700 104 |
| E-mail: tathagata.bandyopadhyay@gmail.com | India |
| | E-mail: rmuk1@hotmail.com |